\documentclass[final, osajnl2,twocolumn,showpacs]{revtex4}
\usepackage{hyperref}
\usepackage{amsmath}
\usepackage{calc}
\usepackage{amsfonts}
\usepackage{amssymb}
\usepackage{graphicx}
\usepackage{bm}
\usepackage{color}
\usepackage{epsfig}
\usepackage{ifthen}

\def\anglea{\alpha}
\def\angleb{\beta}

\begin{document}

\title{Broadband Faraday Isolator}

\author{Micha\l$\,$ Berent $^{1,2,*}$, Andon A. Rangelov$^{1}$ and Nikolay V. Vitanov$^{1}$}

\address{$^1$ Department of Physics, Sofia University, James Bourchier 5 Blvd., 1164 Sofia, Bulgaria}
\address{$^2$ Faculty of Physics, Adam Mickiewicz University, Umultowska 85, 61-614 Pozna\'n, Poland}
\address{$^*$Corresponding author: mberent@amu.edu.pl }

\begin{abstract}
Driving on an analogy with the technique of composite pulses in quantum
physics, we propose a broadband Faraday rotator and thus a broadband optical
isolator, which is composed of sequences of ordinary Faraday rotators and
achromatic quarter-wave plates rotated at the predetermined angles.
\end{abstract}


\maketitle

\section{Introduction}

Optical isolator (optical diode) is a device that allows the light to pass in one direction and blocks it in the opposite direction. It is widely used in laser technology to protect lasers and amplifiers from the detrimental effect of the back reflected light, resulting in the fluctuations of frequency and the output power.

The optical Faraday isolator is made of three parts: an input polarizer (polarized vertically), a $45^{\circ }$ Faraday rotator and an output polarizer (analyzer, polarized at $45^{\circ }$) (see Fig.\ref{rotator}). Light travelling in the forward direction becomes polarized vertically by the input polarizer. The Faraday rotator rotates the polarization by $45^{\circ }$ and the analyzer then enables the light to be transmitted through the isolator.

Light traveling backwards first becomes polarized at $45^{\circ }$ by the analyzer. Then it passes again through the Faraday rotator and, because of its non-reciprocity, the polarization is rotated by another $45^{\circ }$ (in contrast to the quartz rotator, which would rotate the polarization by -$45^{\circ}$). This means that after the second passage the light is polarized horizontally. Since the polarizer is vertically aligned, the light will be extinguished. However, due to the wavelength dependance of the Faraday rotation angle $\theta(\lambda)$, it works efficiently only for a very narrow wavelength window.

Recent advances in broadband (achromatic) retarders \cite{Ardavan,Ivanov}
open the door for the realization of broadband Faraday isolators, which we
present here. We use the analogy between the polarization
Jones vector and the quantum state vector \cite{Kakigi, Seto, Rangelov, Botet, Botet2} to propose a broadband optical isolator, which promises to deliver very high optical isolation and an efficient transmission in an arbitrarily broad range of wavelengths.

\begin{figure}[t]
\centerline{\includegraphics[width=0.8\columnwidth]{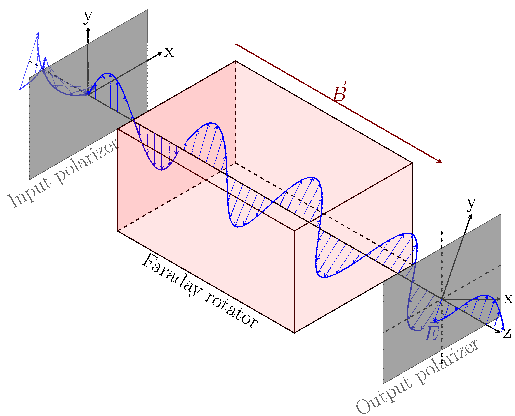}}
\caption{(Color online) Schematic picture of the ordinary Faraday isolator.}
\label{rotator}
\end{figure}

\section{Basics}

Any polarization system can be viewed as a composition of a retarder and a
rotator \cite{Hurvitz}. A rotation at angle $\theta $ in the polarization
plane is described by the following Jones matrix, written in the
horizontal-vertical (HV) basis,
\begin{equation}
\mathfrak{R}(\theta )=\left[
\begin{array}{cc}
\cos \theta & \sin \theta \\
-\sin \theta & \cos \theta%
\end{array}%
\right] .  \label{rotator}
\end{equation}%

A retarder increases the phase of the electric field by $\anglea$ along the fast
axis and retards it by $-\anglea$ along the slow axis, which can be expressed in
the HV basis by the Jones matrix
\begin{equation}
\mathfrak{J}(\anglea)=\left[
\begin{array}{cc}
e^{-i\anglea} & 0 \\
0 & e^{i\anglea}%
\end{array}%
\right] .  \label{retarder}
\end{equation}%
Half- and quarter-wave plates ($\lambda /2$ and $\lambda /4$), are described by $\mathfrak{J}(\pi /2)$ and $\mathfrak{J}(\pi /4)$, respectively.

\begin{figure}[t]
\centerline{\includegraphics[width=0.8\columnwidth]{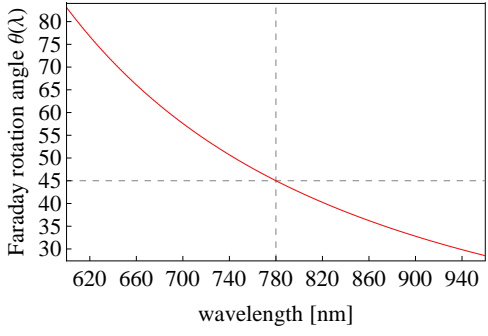}}
\caption{(Color online) Wavelength dependence of the rotation angle of $45^{\circ}$ TGG Faraday rotator tuned to $780$~nm.}
\label{Fig2}
\end{figure}

For the Faraday rotator, the rotation angle is equal to
\begin{eqnarray}
\theta(\lambda) = \nu(\lambda) B L\,,
\end{eqnarray}
where $B$ is the induction of the external magnetic field, $L$ is the length of the Faraday rotator, and $\nu(\lambda)$ is the Verdet constant of the material.
In this paper we will focus on the most popular terbium gallium garnet crystal (TGG); its wavelength dependance can be described as \cite{Yoshida, Villora}
\begin{equation}
\nu(\lambda) = \frac{K}{\lambda_0^2 - \lambda^2}\,,
\end{equation}
where $\lambda_0 = 258{.}2$~nm and $K = 4{.}45\cdot 10^7\,\frac{\text{rad $\cdot$ nm}^2 }{\text{T $\cdot$ m}}$. TGG has optimal material properties for Faraday devices (rotators and isolators) in the range from 400 -- 1100 nm, excluding 470 -- 500 nm (the absorption window).
The dependence of the Faraday rotation angle on the wavelength for a TGG crystal tuned to 780 nm is illustrated in Fig.~\ref{Fig2}.

For most materials the Verdet constant decreases (in absolute value) with increasing wavelength: for TGG it is equal to $-134\,\frac{\text{rad}}{\text{T $\cdot$ m}}$ at $632$~nm and $-40\,\frac{\text{rad}}{\text{T $\cdot$ m}}$ at $1064$~nm.
This is the reason why Faraday isolator works only for a narrow range of wavelengths. However, the recent advances in achromatic wave plates \cite{Ardavan,Ivanov} may be used to make the transmission and isolation of the optical diode insensitive to wavelength.

Let us consider a system of a single Faraday rotator of a rotation angle
$\theta $ sandwiched with two achromatic quarter-wave plates one rotated at
the angle $\anglea $ and the other at $\angleb $ with respect to their fast polarization axes. The Jones matrix $\mathfrak{J}$ for that system has the form
\begin{equation}
\mathfrak{J}_{\theta }(\anglea , \angleb )=\mathfrak{R}(-\angleb)\mathfrak{J}(\pi /4)%
\mathfrak{R}(\angleb)\mathfrak{R}(\theta )\mathfrak{R}(-\anglea )\mathfrak{J}%
(\pi /4)\mathfrak{R}(\anglea ).  \label{jones forward}
\end{equation}%
for forwards direction, whereas for backwards direction it reads
\begin{equation}
\overline{\mathfrak{J}}_{\theta }(\anglea , \angleb )=\mathfrak{R}(-\anglea)\mathfrak{J}^{-1}(\pi /4)%
\mathfrak{R}(\anglea)\mathfrak{R}(\theta )\mathfrak{R}(-\angleb )\mathfrak{J}^{-1}(\pi /4)\mathfrak{R}(\angleb ).  \label{jones backward}
\end{equation}%
In the Faraday magneto-optical effect the external magnetic field induces an anisotropy of the refractive index for the light of left and right circular polarizations. This is why it is more convenient to work in the left-right circular polarization (LR) basis. In this basis the Jones matrix obtains the
form $\mathbf{J}_{\theta }(\anglea , \angleb )=\mathbf{W}^{-1}\mathfrak{J}_{\theta}(\anglea , \angleb )\mathbf{W}$, where $\mathbf{W}$ connects the HV and LR
polarization bases,
\begin{equation}
\mathbf{W}=\tfrac{1}{\sqrt{2}}\left[
\begin{array}{cc}
1 & 1 \\
-i & i%
\end{array}%
\right] .
\end{equation}%
Explicitly, the Jones matrices in the LR basis are
\begin{widetext}
\begin{subequations}
\label{composite element}
\begin{eqnarray}
\mathbf{J}_{\theta }(\anglea, \angleb )&=&\left[
\begin{array}{cc}
-i e^{-i(\anglea-\angleb)}\sin \left( \theta-\anglea+\angleb \right) & -i e^{i(\anglea+\angleb)}\cos \left( \theta-\anglea+\angleb \right)  \\
- i e^{-i(\anglea+\angleb)} \cos \left( \theta-\anglea+\angleb \right) & i e^{i(\anglea-\angleb)}\sin \left( \theta-\anglea+\angleb \right)%
\end{array}%
\right]   \label{composite element forward}\\
\overline{\mathbf{J}}_{\theta }(\anglea, \angleb )&=&\left[
\begin{array}{cc}
-i e^{i(\anglea-\angleb)}\sin \left( \theta+\anglea-\angleb \right) & i e^{i(\anglea+\angleb)}\cos \left( \theta+\anglea-\angleb \right)  \\
i e^{-i(\anglea+\angleb)} \cos \left( \theta+\anglea-\angleb \right) & i e^{-i(\anglea-\angleb)}\sin \left( \theta+\anglea-\angleb \right)%
\end{array}%
\right]   \label{composite element backward}
\end{eqnarray}%
\end{subequations}
\end{widetext}
for forwards and backwards direction, respectively.

\section{Composite Faraday isolator}

Our objective is to construct an effective Faraday isolator that is
operating over a broad range of wavelengths $\lambda $. To this end, we
replace the single composite element described by Eq. \eqref{composite element} with a
sequence of $N$ elements, each with the rotation $\theta_k$ and quarter wave-plates rotated at angles ($\anglea_k$ , $\angleb_k$) with respect to their fast polarization axes. The Jones matrix of such compositions in the LR basis is described by (read from right to left)
\begin{eqnarray}
\mathbf{J}_f^{\left( N\right) } &=& \mathbf{J}_{\theta _{N}}\left(\anglea_N, \angleb
_{N}\right) \mathbf{J}_{\theta _{N-1}}\left(\anglea_{N-1}, \angleb _{N-1}\right) \cdots \nonumber\\
&& \times\mathbf{J}_{\theta _{2}}\left(\anglea_{2}, \angleb _{2}\right)\mathbf{J}_{\theta _{1}}\left(\anglea_{1}, \angleb _{1}\right) \mathbf{.}
\label{overall Jones matrix}
\end{eqnarray}%
for forwards direction and
\begin{eqnarray}
\mathbf{J}_b^{\left( N\right) } &=& \mathbf{\overline{J}}_{\theta _{1}}\left(\anglea_1, \angleb
_{1}\right) \mathbf{\overline{J}}_{\theta _{2}}\left(\anglea_{2}, \angleb _{2}\right) \cdots \nonumber\\
&& \times\mathbf{\overline{J}}_{\theta _{N-1}}\left(\anglea_{N-1}, \angleb _{N-1}\right)\mathbf{\overline{J}}_{\theta _{N}}\left(\anglea_{N}, \angleb _{N}\right) \mathbf{.}
\label{overall Jones matrix}
\end{eqnarray}%
for backwards direction.
We then expand the composite Jones matrices around the assumed rotation angle,
\begin{eqnarray}
		\mathbf{J}_k^{\left( N\right) } (\theta) &=& \mathbf{J}_k^{\left( N\right) }(\theta_0)+\frac{(\theta - \theta_0)}{1 !}\frac{\partial}{\partial \theta} \mathbf{J}_k^{\left( N\right) }(\theta_0)+\hdots \nonumber\\
		&&+ \frac{(\theta - \theta_0)^n}{n !}\frac{\partial^n}{\partial \theta^n} \mathbf{J}_k^{\left( N\right) }(\theta_0)\,,
\end{eqnarray}
where $k = f,\, b$. In the next step we set $\mathbf{J}_f^{(N)}(\theta_0) =\mathbf{J}_b^{(N)}(\theta_0) =\mathbf{J}_{0}$, where $\mathbf{J}_{0}$ is the assumed Jones matrix of the Faraday rotator, and nullify as many derivatives as possible
\begin{eqnarray}
	\frac{\partial^k}{\partial \theta^k} \mathbf{J}_f^{\left( N\right) }(\theta_0) &=& 0 \text{ and }\quad\frac{\partial^k}{\partial \theta^k} \mathbf{J}_b^{\left( N\right) }(\theta_0) = 0\nonumber\\
	&&\text{for }k = 1,2,\dots, \lfloor\frac{N-1}{2}\rfloor.
\end{eqnarray}

The performance of the optical isolators is quantified by its transmission ($T$) and isolation ($D$). Both of these quantities depend on the intensity of light measured after passing the optical diode in the forward and backward direction, respectively
\begin{subequations}
\label{intensities}
\begin{eqnarray}
T = I_{forw}/I_0 &=& \left| \mathbf{P}_{D} \mathbf{J}_f^{\left( N\right) } \mathbf{P}_{V} |in\rangle \right|^2 ,\label{forward intensity}\\
B = I_{back}/I_0 &=& \left| \mathbf{P}_{V} \mathbf{J}_b^{\left( N\right) } \mathbf{P}_{D} |in\rangle \right|^2  ,\label{backward intensity}
\end{eqnarray}
\end{subequations}
where $\mathbf{P}_D$ and $\mathbf{P}_V$ stand for $45^{\circ}$ and vertical polarizers, $| in \rangle$ is the Jones vector for the light entering the isolator. $I_{0}$ has the meaning of the intensity of light at the beginning of the Faraday isolator (we do not discuss the material losses). The isolation is calculated with the formula
\begin{equation}
D = -10 \log_{10}\left[10^{(-L/10)}+B\right]\,,
\end{equation}
where $L$ is the assumed maximum isolation (in dB) and $B$ is given by Eq.~\eqref{backward intensity}.

\subsection{$45^{\circ}$ Faraday rotators}
First, we will show how to construct a broadband $45^{\circ }$ Faraday
rotator $\mathbf{J}_{0}$, which in the LR basis reads (up to a global phase factor)
\begin{equation}
\mathbf{J}_{0}=\tfrac{1}{\sqrt{2}}\left[
\begin{array}{cc}
1-i & 0 \\
0 & 1+i%
\end{array}%
\right] ,  \label{half wave plate Jones matrix}
\end{equation}%
using four composite elements of Eq.~\eqref{composite element} with $\theta = 45^{\circ}$, a rotation angle typically used in commercially available isolators. Numerical calculations revealed that it is impossible to make our composite element to be broadband in both forward and backward directions at once, for a sequences of less than 4 elements. The overall Jones matrix \eqref{overall Jones matrix} for this sequence (in the LR basis) reads
\begin{eqnarray}\label{Faraday-4}
\mathbf{J}_f^{\left( N\right) } &=& \mathbf{J}_{\pi /4}\left(\anglea_4 , \angleb _{4}\right)
\mathbf{J}_{\pi/4 }\left(\anglea_3 , \angleb _{3}\right)\notag\\
&& \times \mathbf{J}_{\pi/4}\left(\anglea_2 ,\angleb _{2}\right) \mathbf{J}_{\pi/4}\left(\anglea_1 ,\angleb _{1}\right).
\end{eqnarray}
The composite matrix in backward direction can be constructed by changing the order of the rotation angles ($\anglea_k$, $\angleb_k$) and replacing the matrices representing quarter-wave plates with their inverses.
We then use rotation angles ($\anglea _{i}$, $\angleb _{i}$) to set $\mathbf{J}_f^{(N)} =\mathbf{J}_b^{(N)} =\mathbf{J}_{0}$ and nullify the first order derivatives of both $\mathbf{J}_f^{(N)}$ and $\mathbf{J}_b^{(N)}$ versus the error in the rotation angle of the Faraday rotator.
We thereby obtain a system of nonlinear algebraic equations for the $8$ rotation angles ($\anglea_{k}$, $\angleb _{k}$) $\left( k=1,2,3, 4\right) $, which we solve numerically. The resulting angles are listed in Table~\ref{tab:angles}.

The sequences of more than four elements (with $45^{\circ}$ Faraday rotators) are also possible, however, they are of less importance as a number of the necessary optical elements will be too large for practical realizations.

\begin{table}
\caption{Rotation angles ($\anglea_k$, $\angleb_k$) (in degrees) for a broadband Faraday rotator with $N=3$ [Eq.~\eqref{Faraday-3}], 4 [Eq.~\eqref{Faraday-4}], and 5 [Eq.~\eqref{Faraday-5}] constituent elements.}
\begin{tabular}{c|c} \hline
 N &  Rotation angles ($\anglea_1$; $\angleb_1$; $\anglea_2$; $\angleb_2$;$\dots$;$\anglea_{N}$;$\angleb_N$) \\
\hline
 3 & (62{.}2; 16{.}6; 175{.}0; 84{.}3; 107{.}7; 152{.}2;) \\
 4 & (59{.}3; 104{.}8; 37{.}3; 171{.}8; 149{.}3; 104{.}8; 37{.}3; 81{.}3) \\
 5 & (175{.}9; 29{.}1; 42{.}9; 74{.}9; 45{.}8; 28{.}2; 15{.}6; 147{.}3; 3{.}5; 85{.}9) \\
\hline
\end{tabular}\label{tab:angles}
\end{table}

\subsection{Arbitrary Faraday rotators}

\begin{figure}[tb]
\centerline{\includegraphics[width=0.8\columnwidth]{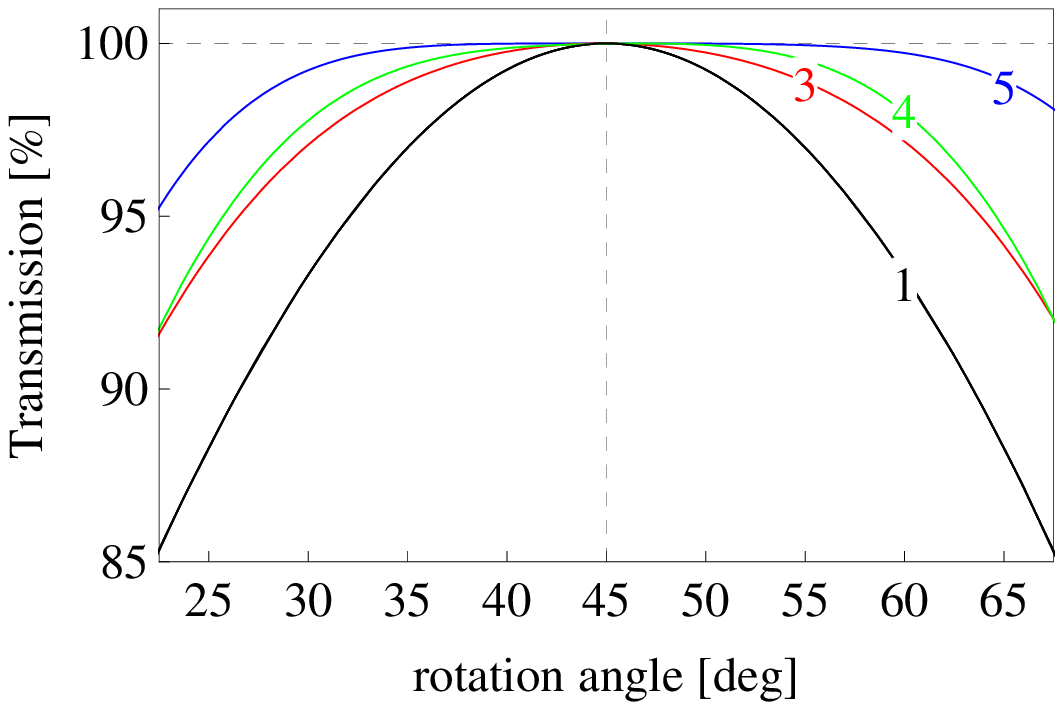}}
\centerline{\includegraphics[width=0.8\columnwidth]{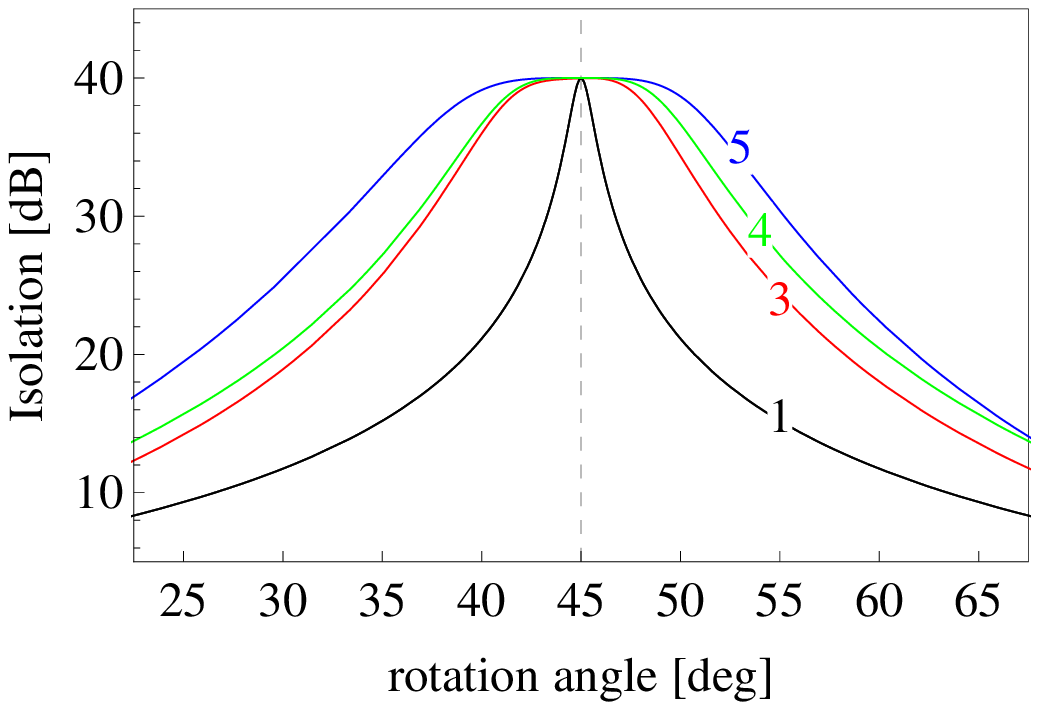}}
\centerline{\includegraphics[width=0.8\columnwidth]{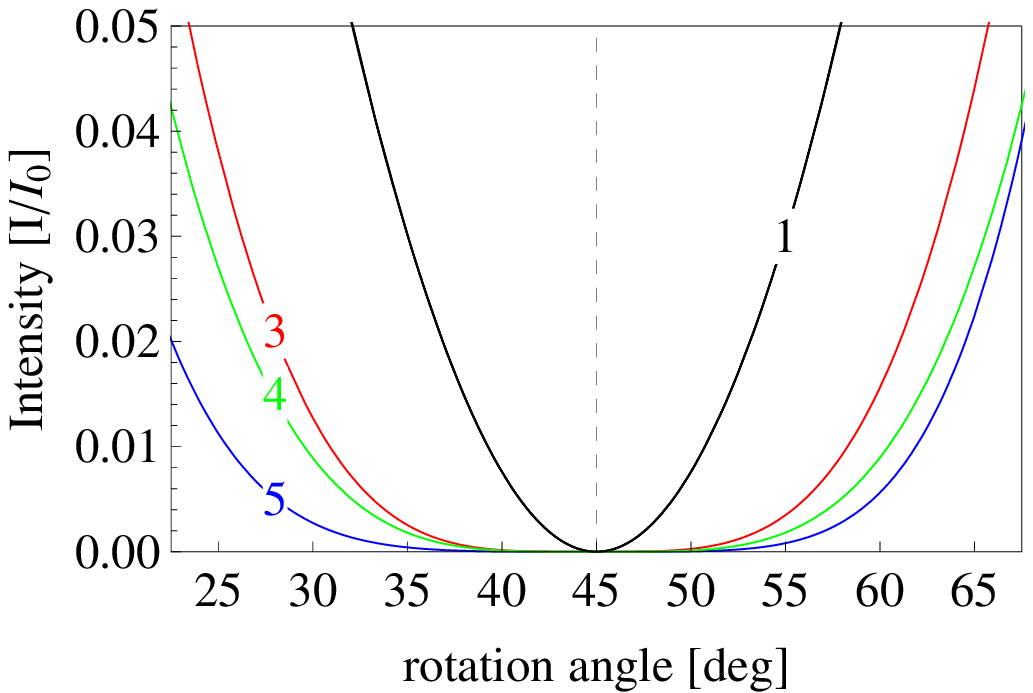}}
\caption{(Color online) Transmission and isolation properties of the composite Faraday isolators as compared to the isolator based on a single rotator (black line), vs the Faraday rotation angle.
The numbers on the curves refer to the sequences of 3 [Eq.~\eqref{Faraday-3}], 4 [Eq.~\eqref{Faraday-4}], and 5 [Eq.~\eqref{Faraday-5}] elements.}
\label{Fig3}
\end{figure}

If one drops the requirement that the Faraday rotation angles must be equal to $45^{\circ}$, then it is possible to construct a broadband isolator composed of only three elements.
We have found that one possible set of rotation angles is then $\pi/4$, $\pi/2$ and $\pi/4$ and thus the composite Jones matrix reads
\begin{equation}\label{Faraday-3}
\mathbf{J}_f^{\left( N\right) } = \mathbf{J}_{\pi/4 }\left(\anglea_3 , \angleb _{3}\right)\mathbf{J}_{\pi/2
}\left(\anglea_2 ,
\angleb _{2}\right) \mathbf{J}_{\pi/4}\left(\anglea_1 ,
\angleb _{1}\right).
\end{equation}

The same can be done with five elements. Now the rotation angles $\theta_k$ are equal $\pi/4$, $\pi/2$, $\pi/2$, $\pi/2$, and $\pi/4$, respectively. The composite Jones matrix in this case has the form
\begin{eqnarray}\label{Faraday-5}
\mathbf{J}_f^{\left( N\right) } &=& \mathbf{J}_{\pi/4 }\left(\anglea_5 , \angleb _{5}\right)\mathbf{J}_{\pi/2 }\left(\anglea_4 , \angleb _{4}\right) \mathbf{J}_{\pi/2 }\left(\anglea_3 , \angleb _{3}\right)\nonumber\\
&& \mathbf{J}_{\pi/2}\left(\anglea_2 ,\angleb _{2}\right) \mathbf{J}_{\pi/4}\left(\anglea_1 ,\angleb_{1}\right).
\end{eqnarray}
The mentioned rotation angles ($\pi/4$ and $\pi/2$) are convenient because such Faraday rotators are commercially available. Nevertheless, the choice of alternative non-standard rotation angles is also possible. The numerically calculated quarter-wave plates rotation angles for the above setups are presented in Table~\ref{tab:angles}.

\begin{figure}[tb]
\centerline{\includegraphics[width=0.8\columnwidth]{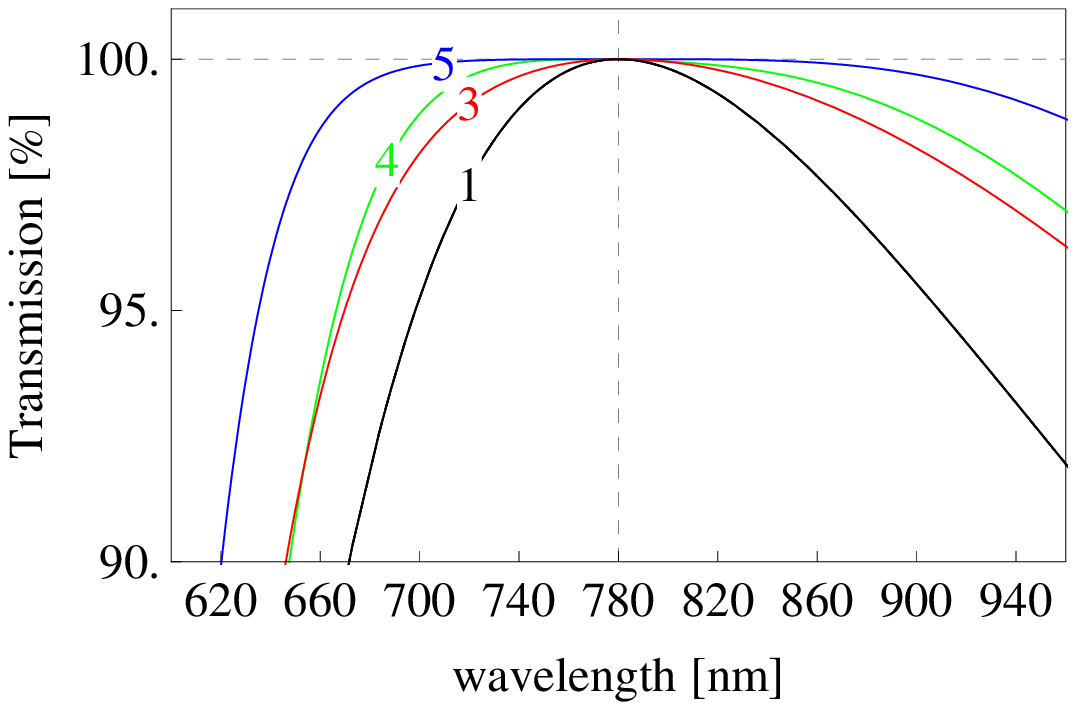}}
\centerline{\includegraphics[width=0.8\columnwidth]{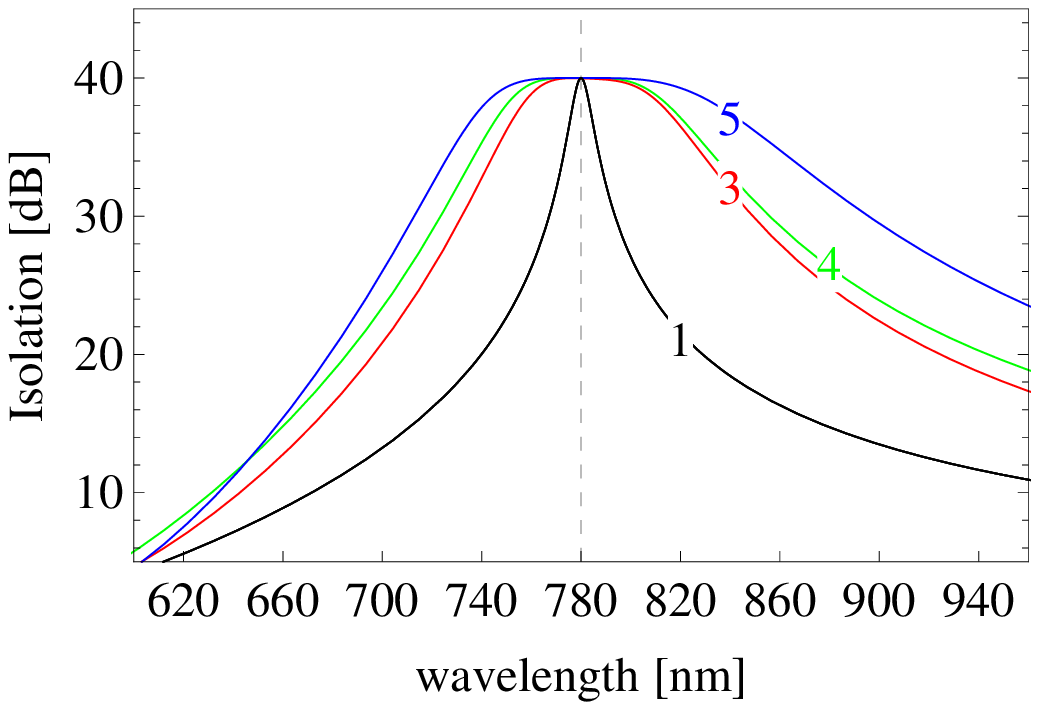}}
\centerline{\includegraphics[width=0.8\columnwidth]{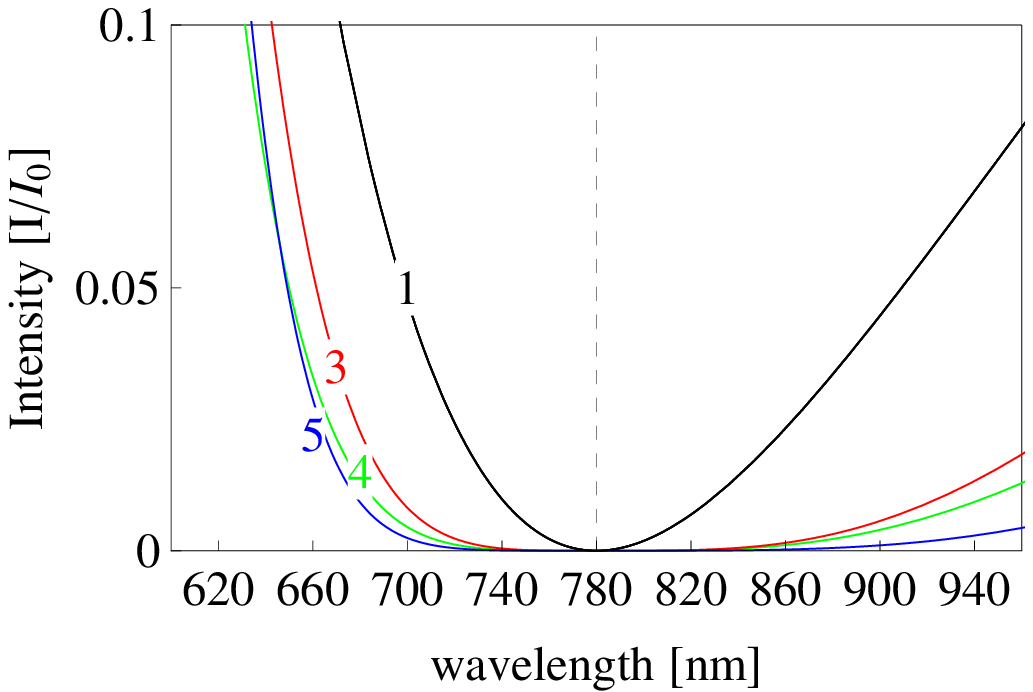}}
\caption{(Color online) The same as Fig.~\ref{Fig3} but vs the wavelength, for narrowband $45^{\circ}$ and $90^{\circ}$ Faraday rotators tuned to $780$~nm.}
\label{Fig4}
\end{figure}

The transmission and isolation profiles for the case of 4 composite elements using four $45^{\circ}$ as well as sequences of 3 and 5 composite elements with $45^{\circ}$ and $90^{\circ}$ Faraday rotators are shown in Figs.~\ref{Fig3} and \ref{Fig4}. One can notice that for all these composite isolators both the transmission and isolation is far more efficient than that of single optical diode. Fig.~\ref{Fig3} shows the performance of the optical elements under study with respect to the rotation angle of the Faraday rotators, whereas in Fig.~\ref{Fig4} the analogous dependance on wavelength is presented. The asymmetry seen in Fig.~\ref{Fig4} stems from the fact that the Faraday rotation angle depends non-linearly on the wavelength (as seen in Fig.~\ref{Fig2}).

One should emphasize that the transmission curves were obtain under the assumption that there are no losses, which is unavoidable in practical realizations. The main cause of the lower transmission are insertion losses. The losses related to the propagation in the optical elements would be of less importance. The mentioned sources of loss can only increase the  isolation properties of the diode as they would mainly result in lower light intensity at the output.

It is important to note that the parameters given in Table~\ref{tab:angles} are not the only possible. As the rotation angles of quarter-wave plates are our control parameters, they allow for an arbitrary manipulation of polarization. The particular choice of the solutions will depend on our requirements.

\section{Conclusions}
We have presented a simple method to construct broadband Faraday isolators with the use of Faraday rotators and quarter-wave plates only,
 with a transmission bandwidth of over 200 nm and an isolation bandwidth of about 50 nm. 
They allow to manipulate polarization by varying the the rotation angle of the quarter-wave plates. The advantage of this isolator is that it is robust against variations in the magnetic field, crystal length and temperature. An experimental implementation with standard optical elements available in most laboratories should be straightforward.

We conclude by pointing out that recently, some suggestions for a non-magnetic optical isolation where proposed \cite{Yu, Kang, Hwang}. It seems that the application of the presented approach should be also possible in those realizations of the optical isolator.

\section*{Acknowledgements}
This work is supported by the European Commission project FASTQUAST and the Bulgarian NSF Grant DMU-03/103.


\begin{thebibliography}{9}

\bibitem{Ardavan}{ A. Ardavan, \textquotedblleft Exploiting the Poincare-Bloch symmetry to design high-fidelity broadband composite linear retarders \textquotedblright , New J. Phys. \textbf{9}, 24 (2007).}

\bibitem{Ivanov}{ S. S. Ivanov, A. A. Rangelov , N. V. Vitanov, T. Peters, and T. Halfmann, \textquotedblleft Highly efficient broadband conversion of light polarization by composite retarders \textquotedblright J. Opt. Soc. Am. A \textbf{29}, 265 (2012).}

\bibitem{Kakigi}{H. Kuratsuji and S. Kakigi, \textquotedblleft Maxwell-Schrodinger equation for polarized light and evolution of the Stokes parameters,\textquotedblright Phys. Rev. Lett. 80, 1888–1891 (1998).}

\bibitem{Seto}{H. Kuratsuji, R. Botet, and R. Seto, \textquotedblleft Electromagnetic gyration,\textquotedblright Prog. Theor. Phys. 117, 195–217 (2007).}

\bibitem{Rangelov}{A. A. Rangelov, U. Gaubatz, and N. V. Vitanov, \textquotedblleft Broadband adiabatic conversion of light polarization,\textquotedblright Opt. Commun. 283, 3891–3894 (2010).}

\bibitem{Botet}{R. Botet and H. Kuratsuji, \textquotedblleft Light-polarization tunneling in optically active media,\textquotedblright J. Phys. A 41, 035301 (2008).}

\bibitem{Botet2}{R. Botet and H. Kuratsuji, \textquotedblleft Stochastic theory of the Stokes parameters in randomly twisted fiber,\textquotedblright Phys.Rev.E81,036602(2010).}

\bibitem{Hurvitz}{ H. Hurvitz and R. C. Jones, \textquotedblleft A new calculus for the treatment of optical systems\textquotedblright , J. Opt. Soc. Am. \textbf{31}, 493-495 (1941).}

\bibitem{Yoshida}{ H. Yoshida, \textquotedblleft Optical properties and Faraday effect of ceramic terbium gallium garnet for a room temperature Faraday rotator \textquotedblright , OPTICS EXPRESS \textbf{19}, 15181 (2011).}

\bibitem{Villora}{ E. G. Villora, \textquotedblleft Faraday rotator properties of {Tb3}[Sc1.95Lu0.05](Al3)O12, a highly transparent terbium-garnet for visible-infrared optical isolators \textquotedblright , Appl. Phys. Lett. \textbf{99}, 011111 (2011).}


\bibitem{Yu}{Zongfu Yu and Shanhui Fan, \textquotedblleft Optical isolation: a non-magnetic approach \textquotedblright , Nature Photon. {\bf 5}, 517 (2011).}

\bibitem{Kang}{M. S. Kang, A. Butsch and P. St. J. Russell, \textquotedblleft Reconfigurable light-driven opto-acoustic isolators
in photonic crystal fibre \textquotedblright , Nature Photon. {\bf 5}, 549-553 (2011).}

\bibitem{Hwang}{J. Hwang, M. H. Song, B. Park, S. Nishimura, T. Toyooka, J. W. Wu, Y. Takanishi, K. Ishikawa, H. Takezoe, \textquotedblleft Electro-tunable optical diode based on photonic bandgap liquid-crystal heterojunctions \textquotedblright, Nature Materials {\bf 4}, 383 (2005).}

\end{thebibliography}
\end{document}